\documentclass[%
reprint,
superscriptaddress,
frontmatterverbose,
 amsmath,amssymb,
prb,
]{revtex4-1}

\usepackage{amsfonts,amsmath,amssymb,amsthm}
\usepackage{dsfont}

\usepackage{graphicx}% Include figure files
\usepackage{dcolumn}% Align table columns on decimal point
\usepackage{bm}% bold math
\usepackage{hyperref}

\hypersetup{
	pdfnewwindow=true,      % links in new window
	colorlinks=true,       % false: boxed links; true: colored links
	linkcolor=blue,          % color of internal links (change box color with linkbordercolor)
	citecolor=blue,        % color of links to bibliography
	filecolor=blue,      % color of file links
	urlcolor=blue,          % color of external links
}

\usepackage{color}
\usepackage{amsmath}
\renewcommand\boldsymbol{\bm}

\usepackage{lineno}
\makeatother
\usepackage{ulem}

\newcommand{\be}{\begin{equation}}
\newcommand{\ee}{\end{equation}}

\def\ket#1{\mathinner{|{#1}\rangle}}
\newcommand{\abo}{{\textit{ab initio}}}
\newcommand{\os}{{OsCl$_3$}}
\newcommand{\ru}{{$\alpha$-RuCl$_3$}}
\newcommand{\jef}{{$j_\text{eff}$}}

\newcommand{\cf}{$\Delta^\text{CF}$}
\newcommand{\tcf}{$\Delta^{\mathrm{CF}}_{\mathrm{tri}}$}

\usepackage{braket}
\usepackage{array}
\usepackage{booktabs}

\newif\ifdraft
\drafttrue
\draftfalse

\begin{document}

% \preprint{APS/123-QED}

\title{Computational exploration of a viable route to the Kitaev-quantum spin liquid phase in monolayer OsCl$_3$}

\author{Qiangqiang Gu}
%\affiliation{International Center for Quantum Materials, School of Physics, Peking University, Beijing 100871, China}
\affiliation{School of Mathematical Science, Peking University, Beijing 100871, China}
\affiliation{AI for Science Institute, Beijing, China}

\author{Shishir Kumar Pandey}\email{shishir.kr.pandey@gmail.com}
\affiliation{AI for Science Institute, Beijing, China}

\author{Yihao Lin}
\affiliation{International Center for Quantum Materials, School of Physics, \\ Peking University, Beijing 100871, China}

\date{\today}% It is always \today, today,

\begin{abstract}
In this computational study, we explore a viable route to access the Kitaev-Quantum Spin Liquid (QSL) state in  recently synthesized monolayer of a so-called spin-orbit assisted Mott insulator OsCl$_3$. In addition to other magnetic ground states in different regions, the small $J_\text{H}$/$U$ region of our Hubbard $U$--Hund's $J_\text{H}$ quantum phase diagram, obtained by combining second-order perturbation and pseudo-Fermion renormalization group calculations, hosts Kitaev-QSL phase. Only Kitaev interaction of a smaller magnitude is obtained in this region. Negligibly small farther neighbor interactions appear as a distinct feature of monolayer \os, suggesting this material to be a better candidate for the exploration of
possible Kitaev-QSL state than earlier proposed materials. Insights from our study might be useful to probe magnetic phase transitions by purportedly manipulating $U$ and $J_\text{H}$ with epitaxial strain in advanced crystal growth techniques. 

%\QG{While the Kiteaev Quantum spin liquid (QSL) has been extensively studied  based on the Kitaev models with effective interactions, the direct relationship between Kitaev physics and the intrinsic Coulomb interactions remains unrevealed. Using the recently synthesized monolayer \os~as a case study, we explore a viable route to access the QSL state in the so-called spin-orbit assisted Mott insulators. By varying Coulomb interaction strength of Hubbard $U$ and Hund's $J_\text{H}$, parameters which purportedly can be tuned in advance crystal growth techniques by epitaxial strain, we obtain a substantial QSL phase in our considered model's magnetic quantum phase diagram. We use second-order perturbation methods and pseudo-Fermion renormalization group calculations to obtain the diagram, which also includes other magnetic phases. Insights from our study may be helpful for designing new Kitaev QSL candidate materials.}

\end{abstract}

%Use showkeys class option if keyword

\maketitle

%\section{Introduction}
The Kitaev honeycomb model with bond-dependent Ising-type exchange interactions on a honeycomb lattice 
naturally hosts the novel quantum spin liquid (QSL) state~\cite{kitaev}. The realization of the Kitaev model in so-called spin-orbit coupling (SOC) assisted Mott insulators was proposed by Jackeli and Khaliullin~\cite{khul1}. This requires SOC active transition-metal ions to be in edge-shared octahedral crystal field (CF) of anions, and Sr$_2$IrO$_4$ was initially proposed as an example~\cite{khul1}. Since then, many materials like \ru~\cite{ru1,ru2,ru3,rucl3_bnds,rucl3_method} and Cobaltates~\cite{co1,co2,co3,co4,co5,co6,co7,co8,co_skp, co10}, other Iridates~\cite{ir1, ir2, ir3,ir4,ir5,ir6,ir7, liiro_cf} are proposed as pertinent Kitaev-QSL candidates~\cite{qsl_mat_pres,nat_rev_qsl}. In these materials, the formation of $j_{\text{eff}}=1/2$ Kramers doublets, which form the low-energy space in the materials, results from the interplay of octahedral CF (\cf) and SOC. 

Although the dominant magnetic interaction between these pseudospin 1/2 states is proposed to be Kitaev ($K$) type~\cite{khul1}, other {\it undesirable} interactions such as isotropic Heisenberg ($J$) and off-diagonal terms ($\Gamma$), as well as farther neighbor couplings, also appear, causing long-range magnetic order in the materials mentioned above, and driving them away from a possible QSL ground state. For example, the appearance of third nearest neighbor (3NN) $J_3$~\cite{rucl3_ref} and strong neutron absorbing Ir ions along with rapidly decaying magnetic form factor with an increase in  momentum transfer in inelastic neutron scattering experiments on Iridates are some of the experimental challenges~\cite{ir4, nat_rev_qsl}. In another example, smaller SOC strength of Ru$^{+3}$ ions as well a significant $J_3$ in \ru~are the hurdles in the quest of QSL state in this material.
%These additional interactions lead to further extensions of the Kitaev model like, $J$-$K$, $J$-$K$-$\Gamma$, $J$-$\Gamma$-$J_3$ and $J$-$K$-$J_3$~\cite{ir1, rucl3_bnds,rucl3_ref, ph_dia1, ph_dia2,ph_dia3,ph_dia4, ph_dia5, ph_dia6}. 
%Reasons behind appearance of these terms might be the direct overlap between TM $d$ orbitals or the additional trigonal/tetragonal (\tcf/\ttcf) distortions of the octahedron~\cite{SCorr_young}.
%causes $d^5$ ($d^7$ for cobaltates) valence state turn into $t_{2g}^5$ ($t_{2g}^5 e_g^2$ in cobaltates) configuration. This means a single-hole configuration
%with effective orbital angular momentum of $l_{\text{eff}} = 1$. The SOC interaction further couples $l_{\text{eff}} = 1$ orbital angular moment with the effective spin $s$ = 1/2 ($s$ = 3/2 for cobaltates) moment
%forming the $j_{\text{eff}}=1/2$ Kramers doublet state~\cite{qsl_mat_pres}. 
Despite such inevitable challenges, attempts have been made to access the QSL state using external knobs. For example, magnetic interactions in \ru~were manipulated by means of the external magnetic field to achieve the QSL state~\cite{mag_field_prl,mag_field_natcom,mag_field_rev}. In another theoretical attempt by Liu {\it et. al.}~\cite{co4}, \tcf~was proposed as a tunable parameter with pressure to realize QSL state in Cobaltates. Towards this goal, our computational study in this article aims to explore the Kitaev-QSL state in SOC-assisted Mott insulators through manipulation of the fundamental electronic parameters $U$ and $J_\text{H}$. Our study is also motivated in part by recent theoretical work that argues for the experimental tuning of $U$ and $J_\text{H}$ parameters in perovskites with epitaxial strain, based on advancements in experimental epitaxial growth techniques~\cite{u_tune_strain}.
%Meaningful insights about the influence of these external knobs on materials property can be obtained computationally by mapping underlying interactions present in a material onto corresponding Hamiltonian terms and exploring the relevant parameter space. The most common and widely acceptable model Hamiltonian used to study these SOC assisted Mott insulators is given in Eq.~\ref{eq:full}.

To realize our goal, we consider the OsCl$_3$ monolayer as a case study. Our material choice is based on the fact it satisfies all the necessary criteria mentioned earlier to host strong Kitaev interactions, $viz$ Os$^{+3}$ 5$d^5$ ions in the edge-shared OsCl$_6$ octahedron possess large SOC interaction. In a recent experimental study by Kataoka {\it et. al.}~\cite{oscl_exp}, magnetization and heat capacity measurements on Os$_{0.81}$Cl$_3$ revealed the absence of any long-range magnetic ordering of Os$^{+3}$ spins arranged on the triangular lattice (with local honeycomb domain formation) up to 0.08 K temperature, which can be an indication of underlying strong magnetic frustration in this compound, also a building block for Kitaev physics. A previous computational attempt to study the topological properties of this compound proposed a ferromagnetic (FM) ground state for the monolayer OsCl$_3$~\cite{oscl_theory}, but a scrupulous investigation of its magnetic properties is warranted in the light of the fact that the iso-electronic \ru~has a zigzag (ZZ) antiferromagnetic (AFM) ground state. From a computational modeling perspective, electronic parameters $U$ and $J_\text{H}$ are part of Hubbard-Kanamori $H_{\text{int}}$ Hamiltonian in Eq.~\ref{eq:full}. The full model Hamiltonian $\mathcal{H}$ in Eq.~\ref{eq:full} has been quite successful and hence, widely accepted for the estimation of magnetic interactions in these materials~~\cite{rucl3_ref,rucl3_method, SCorr_young,co_skp,sr4rho6}. Thus, one can explore the access to the Kitaev-QSL phase by tuning the $U$ and $J_\text{H}$ parameters and calculating the corresponding magnetic interactions which is the objective of this article.
%$H_0 = H_{\text{cf}} + H_{\text{soc}} + H_{\text{int}}$
 
 We first used \abo~calculations of higher accuracy to show that the ZZ and FM configurations in this material are nearly degenerate, indicating highly competing magnetic interactions that may be a reason for the absence of long-range magnetic ordering in this compound. We then varied $U$ and $J_\text{H}/U$ in the second-order perturbation calculations to obtain magnetic interactions and input them into pseudofermion functional renormalization-group (pf-FRG) calculations, which allowed us to obtain a $U$--$J_\text{H}/U$ quantum phase diagram. At small values of $J_\text{H}/U$, we found the emergence of only $K$-type 1NN interaction, albeit of smaller magnitude, resulting in the Kitaev-QSL state occupying a substantial part of the phase diagram. In addition to the QSL state, we also find FM, ZZ, and N\'eel states present in different parts of the phase diagram. Much more advantageous than previously suggested Kitaev-QSL material candidates like Iridates, \ru~and Cobaltates, we find that the second and third neighbor magnetic interactions are negligibly smaller in the monolayer \os, suggesting this material to be a better candidate for the exploration of possible Kitaev-QSL state.
%\section{Magnetic ground state}

{\it Crystal structure and magnetism -- ab initio calculations:} We consider the $C2/m$ honeycomb crystal structure of the monolayer \os~which is similar to \ru~\cite{oscl_theory,oscl_theory2}.  Very recently, a triangular lattice with nanometer-sized honeycomb domains of Os atoms has been observed experimentally~\cite{oscl_exp}. The structure is shown in Fig.~\ref{fig:fig01}(a) and Fig.~S1 of supplementary material (SM)~\cite{sm}. The full optimization of lattice constants within highly accurate \abo~calculations after imposition of both ZZ and FM state brings nearly identical crystal structure in both cases. 
Our phonon calculations with this optimized crystal structure indicate that the honeycomb lattice of \os~monolayer is at least dynamically stable. Details of \abo~calculations, optimized crystal structure, and phonon calculations can be found in Sec.~S1 of SM~\cite{sm}.

\begin{figure}[ht]
\centering
\includegraphics[width=8.0 cm]{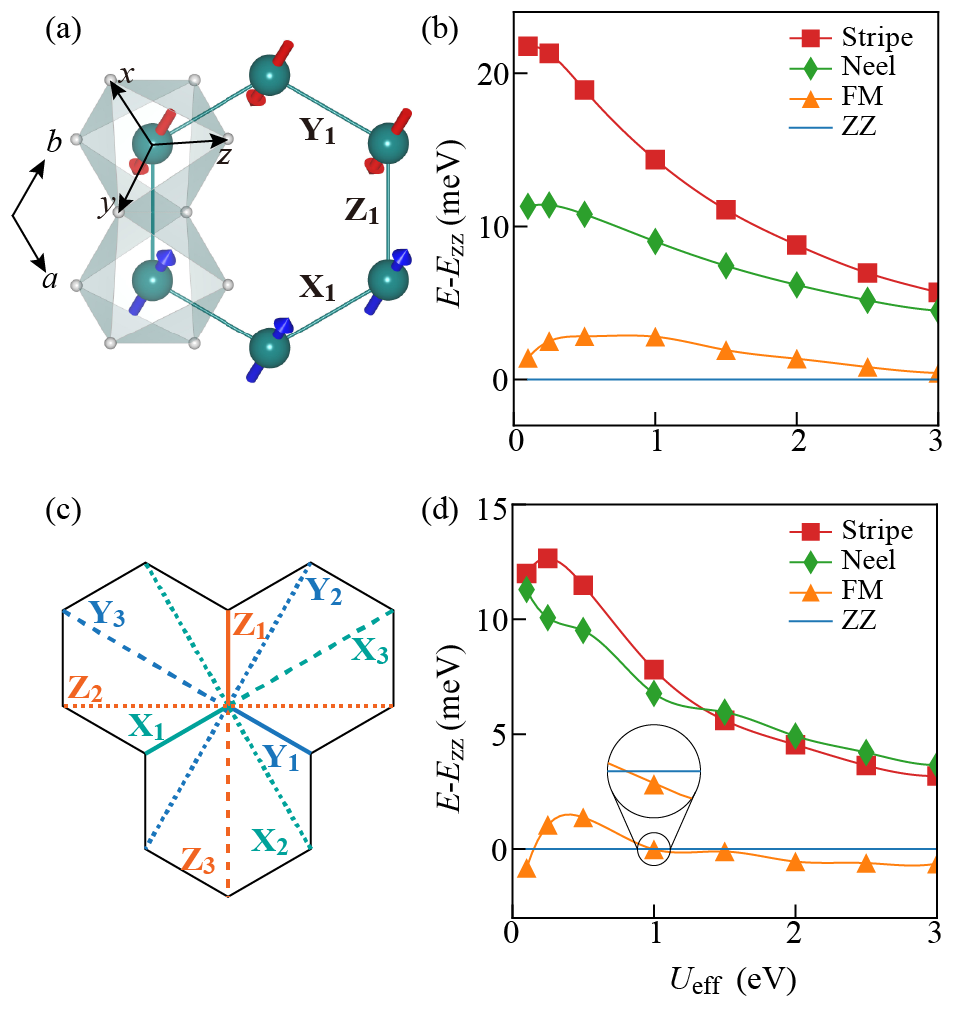}
\caption{(a) Top view of out-of-plane spin arrangement on a hexagon in ZZ magnetic state in monolayer \os. 
Spins point towards the center of one of 
the Cl-Cl edges in an Os-Cl$_6$ octahedron. Big Cyan and small gray balls represent Os and Cl 
atoms; $x$, $y$ and $z$ are local axes while $a$ and $b$ are crystallographic axes. 
 Variation of the total energy of different magnetic configurations relative to ZZ state with 
 $U$ on Os $d$ states for, (b) out-of-plane spin alignment, and (d) alignment of spins along $a$ axis. (c) Color-coded X, Y, and Z bonds up to 3NN are shown.}
  \label{fig:fig01}
\end{figure} 
To understand the nature of magnetism and underlying frustration, highly accurate \abo~total energy calculations are performed, with tighter energy and force convergence criterion, after imposing various magnetic configurations like FM, ZZ, N\'eel, and Stripe. We varied $U$ on Os $d$ states in the range 0.1--3 eV and the SOC effect is self-consistently included. The results, shown in Fig.~\ref{fig:fig01}(b) reveal that the ZZ state is lowest in energy when the spins are canted in an out-of-plane direction towards the center of one of the Cl-Cl edges in an OsCl$_6$ octahedron (see Fig.~\ref{fig:fig01}(a)), with the closely competing FM state higher in energy. Other AFM configurations like N\'eel and stripe are energetically much higher. Such orientation of magnetic moments was also proposed for Na$_2$IrO$_3$~\cite{fengji_ir, ir_mom_nat}. The energy difference between the ZZ and FM state nearly vanishes for the in-plane arrangement of spins along the $a$ axis as shown in Fig.~\ref{fig:fig01}(d). This is consistent with the previous study by Sheng {\it et. al.}~\cite{oscl_theory}. The nearly degenerate FM and ZZ configurations, which differ by a single 1NN spin flip, are indicative of large magnetic frustration present in the system, explaining the absence of long-range magnetic ordering observed in experiment~\cite{oscl_exp}. The larger energy difference between other AFM states, like N\'eel and stripe, and ZZ state may indicate that the exchange interactions in monolayer \os~may not be weak. In the most simplistic second-order perturbation theory, magnetic exchange interactions are $\propto$ $t^2$/$U$ at large $U$ values if one treats hopping $t$ as a perturbation. The expected 1/$U$ behavior~\cite{sarma4d5d, Pandey_2017} of magnetic stability curves in Fig.~\ref{fig:fig01} (b) and (d) at large values $U$  is apparent. We also find that the in-plane magnetic anisotropy for monolayer \os~is quite small (see Fig. S3 of SM~\cite{sm}). Next, we describe the electronic Hamiltonian considered in this study for exploration of $U$--$J_\text{H}$ quantum phase diagram in the quest of QSL state.  
 
 {\it Model Hamiltonian:} The full Hamiltonian for $d^5$ systems is given as~\cite{rucl3_ref,rucl3_method,SCorr_young,co_skp},
 \begin{equation}\label{eq:full}
    \mathcal{H} = H_{\text{hop}} + H_{\text{cf}} + H_{\text{soc}} + H_{\text{int}}
\end{equation}
where terms on the right-hand side are hopping, CF, SOC, and Hubbard-Kanamori Hamiltonians, respectively. For Mott insulators, in limit $t \ll U$, one can separate the ``onsite'' term 
$H^0_i$ from $\mathcal{H}$. $H^0_i$, represented in basis  
$\psi^{\dagger}_{i\sigma}$ = $[d^\dagger_{z^2}$, $d^\dagger_{xz}$, $d^\dagger_{yz}$, $d^\dagger_{x^2-y^2}$, $d^\dagger_{xy}]_{i\sigma}$ for site $i$ and spin $\sigma$, then reads as, 
\begin{eqnarray}
H^0_i &=& H_{\text{cf}} + H_{\text{soc}} + H_{\text{int}} \nonumber \\
&=& \sum_{\sigma} \psi_{i\sigma}^\dagger {\Delta^\text{CF}_i}\psi_{i\sigma} + \sum \lambda \bm L_i \cdot \bm s_i \nonumber \\ 
&+& \frac{U}{2} \displaystyle\sum_{\alpha} n_{i\alpha \sigma} n_{i\alpha \sigma'} 
+ \frac{U'}{2} \displaystyle\sum_{\alpha \ne \beta} n_{i \alpha}n_{i \beta} \nonumber \\ 
&-&\frac{J_\text{H}}{2} \displaystyle\sum_{ \sigma, \sigma', \alpha \ne \beta} \psi^\dagger_{i\alpha \sigma} \psi_{i\alpha \sigma'} \psi^\dagger_{i\beta \sigma'} \psi_{i\beta \sigma} \nonumber \\
 &-& \frac{J'}{2} \displaystyle\sum_{ \sigma \ne \sigma', \alpha \ne \beta} \psi^\dagger_{i\alpha \sigma} \psi_{i\beta \sigma'} \psi^\dagger_{i\alpha \sigma'} \psi_{i\beta \sigma} 
\label{eq:h0}
\end{eqnarray}
 In above expression,  $U$/$U'$ are
intraorbital/interorbital Coulomb interaction terms, and  $J_\text{H}$ and $J'$ are Hund's coupling
and pair hopping interaction,
respectively. $U'$ = $U$ - 2$J_\text{H}$ and $J_{\text H}$ = $J'$ is considered in the above equation. To bring complete rotational invariance of $H^0_i$ in the presence of $e_g$ orbitals, the inclusion of additional three and four orbitals terms in $H_\text{int}$ should be considered. However, for large $\Delta^\text{CF}_i$, such additional terms have an imperceptible influence on magnetic interactions. A discussion around this point is included in Sec. 4(a) of SM~\cite{sm}.
$\Delta^\text{CF}_i$ and  hopping amplitudes $t_{ij}$ in $H_{\text{hop}}$ = $\sum_{i\neq j,\sigma} \psi_{i\sigma}^\dagger T_{ij}\psi_{j\sigma}$ are obtained from a Wannier based tight-binding (TB) model excluding SOC effect. The strength of  SOC in monolayer \os, $\lambda$ = 0.355 eV, is estimated by a band-structure fitting procedure (see Sec S2~\cite{sm} for details). As proposed earlier, it might be possible to experimentally tune parameters $U$ and $J_\text{H}$ by epitaxial strain in advance growth techniques~\cite{u_tune_strain}. Hence, we vary these two parameters to explore the possibility of QSL state in monolayer \os. For a $d^5$ system, the dimension of Hilbert spaces for $H^0_i$ is $C_{5}^{10}=$ 252 and by exactly diagonalizing it, we obtain the lowest two eigenstates as the \jef~=1/2 Kramers doublet. We then use the second-order perturbation method, treating hopping as a perturbation, to estimate magnetic interactions between these pseudo-spin states. Details of the methodology and discussions on the validity of low-energy \jef~=1/2 picture and atomic features which can be observed in experiments for monolayer \os~are presented in Sec. S3, S4 and S6 of SM~\cite{sm}.

\begin{figure*}[ht]
\centering
\includegraphics[width=14.0 cm]{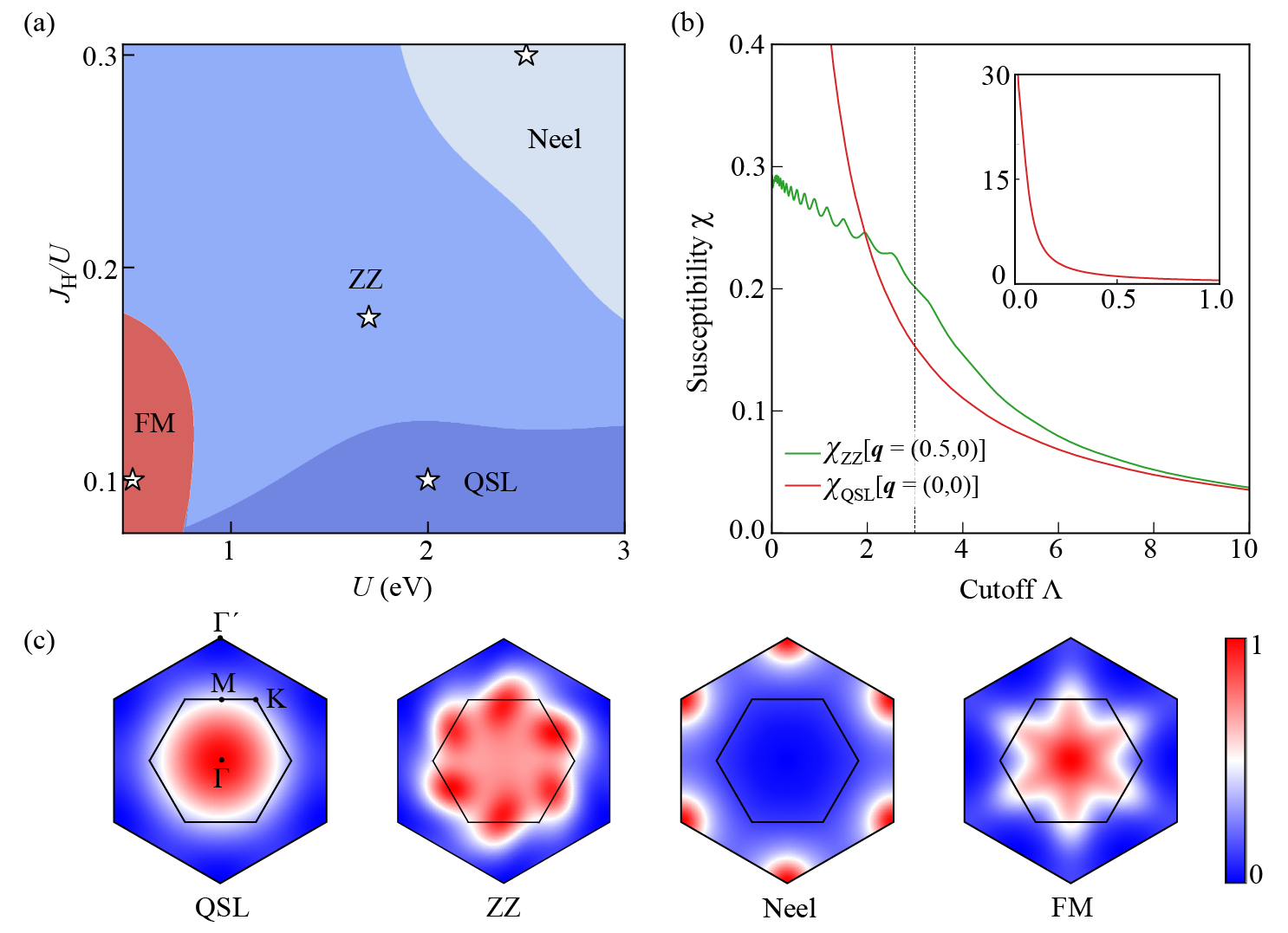}
\caption{(a) Magnetic quantum phase diagram in $U$--$J_\text{H}$/$U$ parameter space obtained using a combination of second-order perturbation and pf-FRG calculations. The range considered is $U$ = 0.5 -- 3.0 eV and $J_\text{H}$/$U$ = 0.075 -- 0.3 are considered. Obtained magnetic phases are FM, ZZ, N\'eel, and QSL magnetic states. Static spin correlation plots in (c) correspond to the points depicted by star symbols in (a). Intensity is normalized for each plot in (c), and the inner and outer hexagons in these plots denote the first and extended BZ of monolayer \os. (b) The renormalization group flow of the magnetic correlations ($\chi$) for the two cases, ZZ (green curve) and QSL (red) states with an inset showing the flow at low frequencies for the QSL state. Vertical dash lines show the break of the flow of $\chi$ for the ZZ state while its smooth flow for the QSL state is indicative of no spontaneous symmetry breaking.
}
  \label{fig:fig02}
\end{figure*}

 {\it Magnetic interactions and quantum phase diagram:} Estimated magnetic interactions up to 3NN (Fig.~\ref{fig:fig01}(c) depicts bonds up to 3NN) from second-order perturbation method are used as input for the pf-FRG calculations. We then analyzed the ground states by plotting the static spin correlation function, obtained from pf-FRG calculations, in the first and extended BZ of monolayer \os. Our choice of the pf-FRG method as a quantum toolkit is based on its earlier demonstrated success in exploring quantum phase diagrams on different lattice and spin models~\cite{pffrg1,*pffrg2,*pffrg3,pffrg4,*pffrg5,*pffrg6}.
 Our obtained quantum phase diagram for a range of $U$ (0.5 - 3.0 eV) and $J_\text{H}$/$U$ (0.075 - 0.3) is shown in Fig.~\ref{fig:fig02}(a). This range of parameters is chosen so that the phase space is extended on either side of the point $U$ = 1.7 eV and $J_\text{H}$ = 0.3 eV ($J_\text{H}$/$U$ $\sim$ 0.176) previously considered for 5$d$ system such as Iridates in the literature~\cite{rucl3_ref}. Details of pf-FRG calculations are given in Sec.~S5 of SM~\cite{sm}. We will now discuss this phase diagram in detail.

\begin{table}[!ht]
 \centering
 \caption{ Estimated bond-dependent magnetic interactions for monolayer \os~using second-order perturbation method at $U$ = 1.7 eV, $J_{\text{H}}$ = 0.3 eV and $\lambda$ = 0.355 eV.
 Values of  Heisenberg $J$, Kitaev $K$, 
 off-diagonal $\Gamma$, $\Gamma'$ and asymmetric DMI $\bm{D}$ terms along with diagonal and off-diagonal 
 anisotropic terms $\xi$ and $\zeta$ are listed. 3NN interactions as well as some 2NN 
 interactions are $<$ 0.05 meV and hence not listed here.}

\begin{tabular}{ccccccc}
\hline \hline
%\specialrule{0em}{0pt}{2pt}
% &  & \multicolumn{3}{c}{Set 1}  &   \\
%\specialrule{0em}{0pt}{2pt}
%\cline{2-7}
%\specialrule{0em}{0pt}{2pt}
%  Bond             & $J$        & $K$     & $\Gamma$  & $\Gamma'$  & $\xi$  & $\zeta$  \\
% \hline 
%\specialrule{0em}{0pt}{3pt}
%X$_1$              &   ~~0.133    & -1.824  &   -0.445  & ~0.078     & ~0.104 & -0.174 \\ 
%Y$_1$              &   ~~0.131    & -1.753  &   -0.449  & -0.104     & -0.062 & -0.228 \\  
%Z$_1$              &   ~~0.100    & -1.755  &   ~0.410  & ~0.308     & -0.100 & ~0.040 \\ 
%X$_2$/Y$_2$/Z$_2$  &   ~-0.211    & ~0.630  &     --    &  --        &   --   &  --     \\
%\specialrule{0em}{0pt}{3pt}
%\cline{2-7} 
%\specialrule{0em}{0pt}{2pt}
%&  & \multicolumn{3}{c}{$\bm{D}_{ij}$}  &   \\
%\hline
%\specialrule{0em}{0pt}{3pt}
%X$_2$           &   &   (-0.022, & -0.044, &  ~0.041)  &   &  \\ 
%Y$_2$           &   &   ( 0.043, & -0.044, & -0.019)  &  &  \\  
%Z$_2$           &   &   ( 0.039, & -0.016, & ~0.047)   &  & \\
%\specialrule{0em}{0pt}{3pt}
%%\cline{2-7}
%%X$_3$/Y$_3$/Z$_3$ & &    &   &   $\le$ $|$0.05$|$  &   &   &     \\ \\
%\hline \hline
\specialrule{0em}{0pt}{2pt}
& &\multicolumn{4}{c}{Magnetic interactions (meV)}  &  \\
\specialrule{0em}{0pt}{2pt}
\cline{2-7}
\specialrule{0em}{0pt}{2pt}
  Bond   & $J$   & $K$ & $\Gamma$ & $\Gamma'$  & $\xi$  & $\zeta$  \\
\hline 
\specialrule{0em}{0pt}{3pt}
X$_1$ &  -1.055    & -14.600  &  ~5.662  &  -3.578 &  ~0.520 & ~2.515    \\ 
Y$_1$ &  -0.961    & -17.215  &  ~6.010  & -0.748 & -0.213 & -4.320 \\  
Z$_1$ &  -0.844    & -15.127  &   ~4.749  & -2.342 & -1.285 &  -4.567 \\
X$_2$ & -0.294     &   ~~0.533  &  ~0.206  & -0.197 &  -- & -0.082     \\
Y$_2$ &  ~0.145    &   ~~0.878  &  ~0.050  &  -0.150 & -- & -0.167     \\
Z$_2$ & -0.258     &  ~~0.453  &  ~0.086  &  -0.246 & -- & -0.050     \\

\specialrule{0em}{0pt}{2pt}
\cline{2-7} 
\specialrule{0em}{0pt}{2pt}
&  &  & \multicolumn{3}{c}{$\bm{D}_{ij}$}     \\
\hline
\specialrule{0em}{0pt}{3pt}
X$_2$           &   & &  ( 0, & 0, &  0)    &  \\ 
Y$_2$           &   & &  ( -0.350, &  0, &  -0.216)   &  \\  
Z$_2$           &   & &  ( -0.108, &  -0.125, & 0 )    & \\
\specialrule{0em}{0pt}{2pt}
\hline \hline
\end{tabular}
\label{tab:mint}
\end{table}

Various magnetic phases $viz$ FM, ZZ and N\'eel and QSL states occupy different parts of the phase diagram. At smaller values of $U$ = 0.5 eV and $J_\text{H}$/$U$ = 0.075 - 0.15, there is a ($<$ 1.5 meV) 1NN FM $J$ which is slightly larger than the AFM $K$ coupling resulting in an FM ground state with a competing ZZ phase. This behavior manifested in a star-like shape of static spin susceptibility (hereafter only susceptibility) plot in Fig.~\ref{fig:fig02}(c) (last plot). Consistent with the FM phase, the major contribution to the susceptibility comes from the zone center while small spreads along $M$-points in the first BZ indicate a competing ZZ AFM phase. Transition to later happens when both couplings change their respective signs with FM $K$ $>$ AFM $J$, which is consistence with a recent observation on Cobaltates~\cite{co_skp}.  At a typical phase point $U$ = 0.5 eV and $J_\text{H}$/$U$ = 0.1 in this region, depicted by white star in Fig.~\ref{fig:fig02}(a), we obtained $J$ = -0.5 meV $K$ = 0.35 meV.

In ZZ region, away from the FM-ZZ phase boundary, other terms like $\Gamma$, $\Gamma'$, $\xi$, $\zeta$ as well as smaller farther neighbor interactions started to emerge in the ZZ region of the phase diagram. However, 1NN dominant $K$ and $J$ terms remain FM and AFM, respectively. We emphasize here that 2NN and 3NN interactions are at least 2 $\sim$ 3 orders of magnitude smaller than 1NN and have no qualitative effect on the magnetic ground state. Values of $U$ = 1.7 eV and $J_\text{H}$ = 0.3 eV bring ZZ state (white star in ZZ region in Fig.~\ref{fig:fig02}(a)) and estimated interactions are listed in Table~\ref{tab:mint}. The corresponding susceptibility plot in Fig.~\ref{fig:fig02}(c) (second from the left) has dominant contributions from $M$-points of the first BZ, consistent with the underlying ZZ magnetic order. In the absence of inversion symmetry on the 2NN bonds, a small Dzyaloshinskii–Moriya interaction (DMI) appears for these magnetic neighbors. We would like to mention that the value of $U$ and $J_\text{H}$ obtained from cRPA calculations also brings a ZZ state, details of which are provided in Sec. S7 of SM~\cite{sm}. ZZ phase occupying the largest part of the phase diagram quantitatively differs in different regions and a related discussion is provided in Sec. S7 of SM~\cite{sm}. Further increase of $U$ ($\approx$ 2 -- 3 eV) and $J_\text{H}$/$U$ ($\approx$ 0.2 -- 0.3) stabilizes N\'eel AFM state, with 1NN AFM $J$ being the dominant interaction while all other interactions are highly suppressed. For a typical phase point $U$ = 2.5 eV, $J_\text{H}$/$U$ = 0.3 (marked by a white star in N\'eel region), 1NN, 2NN, and 3NN $J$ values are 3.39, 0.39 and 0.03 meV, respectively. Contribution from $\Gamma'$ points of the extended BZ in the corresponding susceptibility plot shown in Fig.~\ref{fig:fig02}(c) (third from the left) is consistent with the underlying N\'eel magnetic order. 

Most importantly, a substantial region of the phase diagram, in the range of $U$ = 1.0 - 3.0 eV, $J_\text{H}$/$U$  = 0.075 - 0.1, is occupied by the Kitaev-QSL state. In this region, we only obtained a small but finite 1NN FM $K$ term ($\sim$ 1 meV) in our second-order perturbation calculations. The susceptibility plot for this state shown in Fig.~\ref{fig:fig02}(c) (leftmost) has a contribution from the whole first BZ, with the highest intensity appearing at the $\Gamma$ point. No long-range magnetic ordering and hence, no spontaneous symmetry breaking is expected for this state even at low temperatures. To verify this, we can plot the renormalization group flow of the magnetic correlations ($\chi$) as a function of the renormalization group cutoff ($\Lambda$). Spontaneous symmetry breaking by the onset of a long-range magnetic ordering at a $\Lambda$ results in a kink or break of $\chi$. The smooth evolution of $\chi$ down to the lowest $\Lambda$ can be seen for the QSL state from the red curve in Fig.~\ref{fig:fig02}(b) (see also the inset). For comparison, we also show a similar plot for the ZZ order (green curve) corresponding to interactions listed in Table~\ref{tab:mint}. Clearly, $\chi$, in this case, displays a breakdown at $\Lambda_c$ $\approx$ 3.3 signaling the onset of the ZZ order. 
 
Several remarks are in order. First, though off-diagonal magnetic coupling (e.g. $\Gamma$) is large compared to Iridates~\cite{rucl3_ref}, orders of magnitude smaller farther neighbor magnetic interactions (3NN interactions are all $<$ $|$0.05$|$) distinguishes monolayer \os~these earlier proposed Kitaev-QSL candidates including Iridates~\cite{rucl3_ref, co_skp}. 
 Second, strain as an external knob, proposed in Ref.~\cite{u_tune_strain} for tuning $U$ and $J_\text{H}$ can also be 
 utilized to tune trigonal distortions as suggested in Ref.~\cite{co4}. The strain in this case can also alter 
 the hopping amplitudes between $d$ orbitals. Hence, it would be interesting to probe the role of strain in tuning 
 the magnetic properties of these QSL candidate materials. Third, tuning of $U$ and $J_\text{H}$ to access the QSL region in the phase diagram can also be achieved through the choice of suitable elements from the periodic table. Advancement in material synthesis techniques can be of great help here.  

%($U$ = 3 eV, $J_\text{H}$/$U$ = 0.2 and $U$ = 2 - 3 eV $J_\text{H}$/$U$ = 0.2)
In conclusion, considering recently synthesized monolayer \os~as an example, we have explored a route to access Kitaev QSL state by varying electronic parameters Hubbard $U$ and Hund's $J_\text{H}$. These two parameters purportedly can be tuned experimentally with epitaxial strain. Using highly accurate \abo~total energy calculations, we have shown that the ZZ and FM are nearly degenerate indicating the large magnetic frustration which might be the possible reason behind the absence of long-range magnetic ordering in this material. Combining second-order perturbation and pf-FRG calculations, we presented a magnetic quantum phase diagram in $U$-$J_\text{H}$/$U$ space. A substantial part of this phase diagram hosts the Kitaev QSL state, stabilized by 1NN Kitaev interaction only, with other magnetic phases like FM, ZZ, and N\'eel AFM states
also appearing in different regions of the phase diagram stabilized by changing dominant magnetic interactions. Insights obtained from our study may be helpful in designing experiments on pertinent Kitaev-QSL candidate materials.

\begin{acknowledgments}
QG and SKP contributed equally to this work. SKP gratefully acknowledges valuable feedback from 
Prof. D. D. Sarma. SKP also acknowledges the careful reading of the manuscript and correspondence on magnetic exchange couplings with Ramesh Dhakal and Dr. Stephen M. Winter. 
%We gratefully thank Prof. Ji Feng for providing computation resources.

\end{acknowledgments}

%\bibliography{ref.bib}


\clearpage
\appendix
\renewcommand{\thefigure}{S\arabic{figure}}
\renewcommand{\thetable}{S\arabic{table}}
\setcounter{figure}{0}
\setcounter{table}{0}
\renewcommand{\theequation}{S\arabic{equation}}
\setcounter{equation}{0}
\section*{Supplementary materials}
\subsection{\textit{Ab initio} calculations}
\subsubsection{Details of \textit{ab initio} calculations}
\begin{figure}[ht]
\centering
\includegraphics[width=7.0 cm]{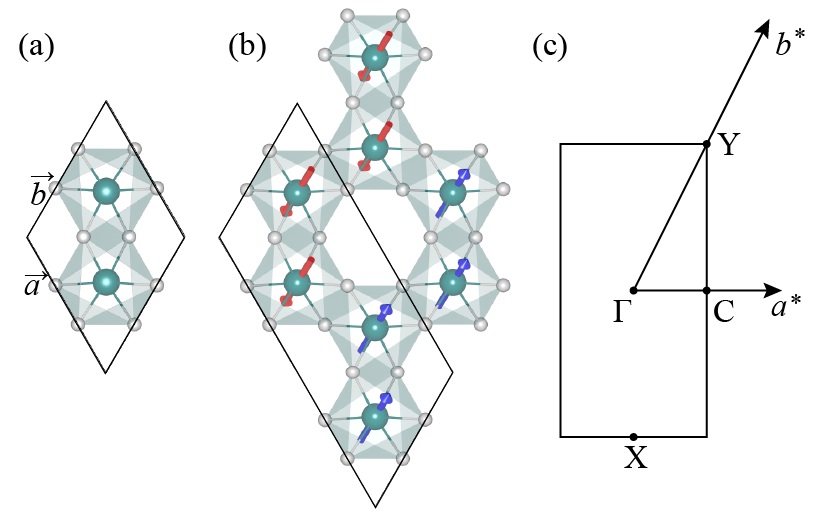}
\caption{(a) Primitive unit cell of OsCl$_3$, (b) the $2\times1\times1$ supercell hosting ZZ magnetic order, and (c) BZ of the ZZ supercell.}
\label{fig:strBZ}
\end{figure}

Density-functional theory (DFT) calculations have been performed using projector-augmented wave method \cite{PAW,PAWpotentials1}, implemented within Vienna \textit{ab initio} simulation package (VASP)~\cite{vaspKresse1993, *vaspKresse1996, *vaspKresse1996_2}.
The Perdew-Burke-Ernzerhof functional \cite{PBE} is used for the exchange-correlation functional within the GGA formalism.
We start with the monoclinic ($C2$/$m$) crystal structure for the monolayer OsCl$_3$ with 2 Os atom unit cell as shown in Fig.~\ref{fig:strBZ}(a).
Full structural relaxation is performed on the $2\times 1 \times 1$ magnetic supercell with FM and ZZ magnetic orders (shown in Fig.~\ref{fig:strBZ}(b)). The energy cutoff for a plane-wave basis is set to 500 eV and $\Gamma$-centered 
$k$-mesh of  $4\times8\times1$ is considered in our calculations. High accuracy of our calculations is ensured by 
 tighter relaxation convergence criterion wherein tolerance considered for total energy is 10$^{-8}$ eV while for force 
 it is $5 \times 10^{-6}$ eV/\AA{}. Full optimization of the crystal structure is performed after considering $U$ = 1 eV under  Dudarev~\cite{dudarev} scheme and spin-orbit coupling (SOC) effect at the self-consistent level. For both the cases of 
 ZZ and FM states, optimization resulted in a similar crystal structure and optimized lattice constants are 
 $a=b=6.1787$ \AA, vacuum $c$ = 25 \AA{}, and $\alpha$ = $\beta$ = 90$^\circ$ and $\gamma \approx$ 120$^\circ$ and atomic coordinates of  the final structure are listed in the Table~\ref{tab:unitcell}.
The primitive unit cell is shown in Fig.~\ref{fig:strBZ} (a) and the Brillouin zone (BZ) for the zigzag (ZZ) supercell is shown in Fig.~\ref{fig:strBZ} (c). 
\begin{table}[!ht]
\centering
\tabcolsep=0.5 cm
\caption{Atomic coordinates in magnetic unit cell of monolayer OsCl$_3$ with space group  $C2/m$. Lattice constants are $a=b=6.1787$ \AA, $c=25$ \AA, $\alpha=\beta=90^\circ$ and $\gamma=119.88^\circ$}
\begin{tabular}{cccc}
\hline \hline 
%\specialrule{0em}{0pt}{2pt}
atom & $x/\boldsymbol{a}$ & $y/\boldsymbol{b}$ & $z/\boldsymbol{c}$ \\
%\specialrule{0em}{0pt}{2pt}
\hline
Os1 & 0.166797 &   0.833203  & 0.500000 \\
Os2 & 0.833203 &   0.166797  & 0.500000 \\
Cl1 & 0.855480 &   0.855480  & 0.446984 \\
Cl2 & 0.144520 &   0.144520  & 0.553016 \\
Cl3 & 0.499663 &   0.143222  & 0.446932 \\
Cl4 & 0.856778 &   0.500337  & 0.553068 \\
Cl5 & 0.500337 &   0.856778  & 0.553068 \\
Cl6 & 0.143222 &   0.499663  & 0.446932 \\
\specialrule{0em}{0pt}{3pt}
\hline \hline
\end{tabular}
\label{tab:unitcell}
\end{table}

\subsubsection{Phonon dispersion} 
\begin{figure}[ht]
\centering
\includegraphics[width=8.0 cm]{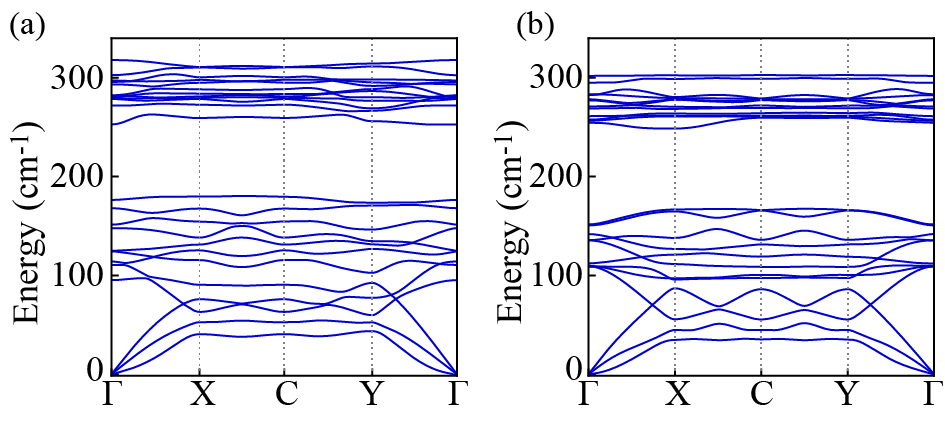}
\caption{Phonon dispersion plots for monolayer OsCl$_3$ with ZZ magnetic order and $U$ = 1 eV on Os $d$ orbitals calculated without (a) and with (b) SOC effect.}
\label{fig:phonon}
\end{figure}

The pristine monolayer OsCl$_3$ has not been observed experimentally yet. As explained in the main text, triangular lattice of Os$^{+3}$ ions with nanometer-sized honeycomb domains is observed recently in experiments~\cite{oscl_exp}
This raises a natural question about the structural stability of the resulting monolayer honeycomb lattice. To answer this, we perform phonon calculations on a 2$\times$4$\times$1 supercell, while considering 
Hubbard $U$ = 1 eV on Os $d$ orbitals and imposing ZZ magnetic configuration. We used the finite displacement method to calculate the force constants and phonon dispersions obtained with and without SOC effects are shown in Fig.~\ref{fig:phonon}. In both cases, crystal structure is at least dynamically stable as no soft modes are found in our calculations. The post-processing of these phonon calculations is done by Phonopy package~\cite{phonopy1,*phonopy2}.

\subsubsection{In-plane magnetic anisotropy}
\begin{figure}[ht!]
\centering
\includegraphics[width=6.0 cm]{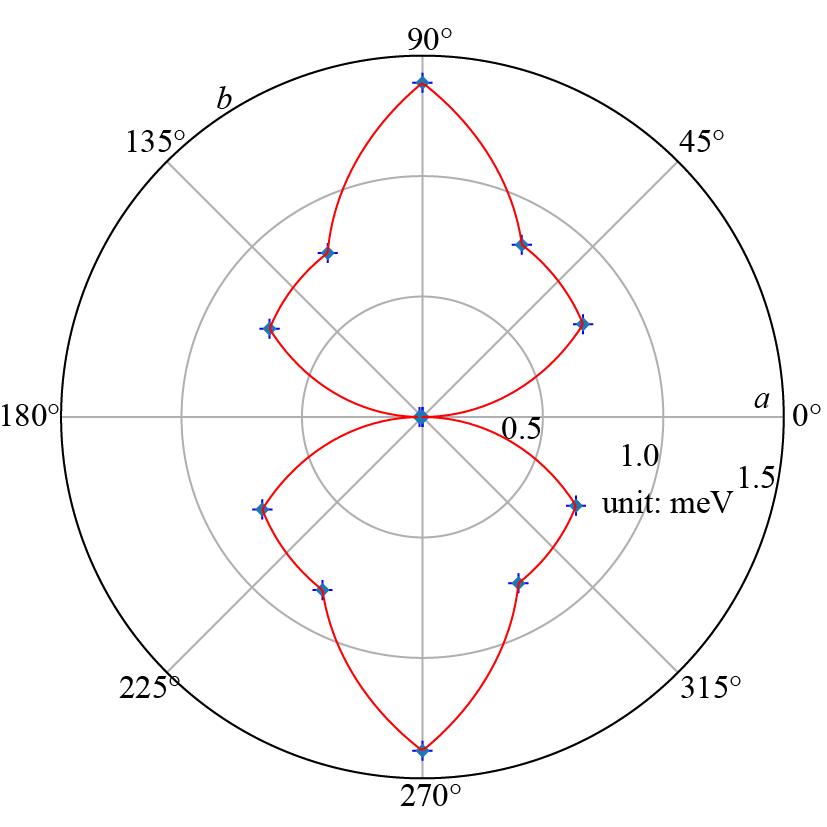}
\caption{In-plane magnetic anisotropy in monolayer OsCl$_3$ estimated considering FM state. While Os spins prefer to 
align along $\bm{a}$, in-plane magnetic anisotropy is small in this material.}
\label{fig:fmaniso}
\end{figure}

To obtain the magnetic anisotropy in monolayer \os, we start with magnetic moments aligned along $\bm{a}$ axis in the FM state. We then rotate these moments by 30$^\circ$ in the $\bm{ab}$ plane and perform total energy calculations. The upper limit of rotation was fixed at 180$^\circ$. Obviously, at 120$^\circ$, magnetic moments are aligned along the $\bm{b}$ axis. The obtained plot is shown in Fig.~\ref{fig:fmaniso}. From this plot, one can conclude that while Os spins prefer to align along $\bm{a}$ axis, the in-plane anisotropic energy is small in monolayer \os.   

\subsection{Estimation of electronic parameters}

\subsubsection{Crystal field splitting and hopping amplitudes}

Crystal field matrix and hopping amplitudes are estimated from a non-spin-polarised, SOC excluded tight-binding (TB) Hamiltonian ($H_\text{TB}$) expressed in the local axes framework depicted in Fig. 1(a). These two terms can be given as, 
$$
H_{\text{TB}} = H_{\text{cf}} + H_{\text{hop}} 
$$
The above Hamiltonian can be obtained by projecting the \textit{ab initio} band structure onto Wannier functions in which all the five Os-$d$ orbitals form the basis~\cite{wannier90,*Franchini2012}.
The choice of the local axes (\bm{$x$}, \bm{$y$}, \bm{$z$}) in Fig. 1(a) along the oxygen atoms is such that  \bm{$c$} = \bm{$x$} + \bm{$y$} + \bm{$z$} lies along trigonal distortion axis of the OsCl$_6$ octahedra. Close agreement between the thus obtained TB model and \abo{} band structure is shown in Fig.~\ref{fig:socfit}(a). Crystal field matrix on a site $i$ ($\Delta^\text{\tiny CF}_i$) in the above term $H_\text{cf}$ is extracted from the onsite part of $H_\text{TB}$ and is given by,

\begin{equation}\footnotesize
	\label{eq:cf}
	\Delta^\text{CF}_i =
	\begin{bmatrix}
		~~2.6322 & -0.0772 & ~~0.2578 & -0.0122 & ~0.0369 \\
		-0.0772  &   ~~0.0479 &  -0.0268  &  ~~0.0709   & ~0.0272 \\
		~~0.2578  &   -0.0268  &  ~~0.0707 &  ~~0.1026   &  ~0.0366 \\
		-0.0122 &   ~~0.0709   &  ~~0.1026   &  ~~2.6244   &  ~0.2800  \\ 
		~~0.0369  &   ~~0.0272 &    ~~0.0366   &   ~~0.2800  &   ~0.0748 \\
	\end{bmatrix}
\end{equation}
Entries in the above matrix are in eV. Hopping amplitudes up to third neighbor are listed below in Table~\ref{tab:hopp}.

\begin{table*}[!ht]
\caption{First, second and third neighbor Os-Os hopping amplitudes (in eV) on different types of bonds listed in the basis $\psi^{\dagger}$ = $[d^\dagger_{z^2}$, $d^\dagger_{xz}$, $d^\dagger_{yz}$, $d^\dagger_{x^2-y^2}$, $d^\dagger_{xy}]$}.
	\begin{tabular}{ccc}
		X$_1$-bond & Y$_1$-bond & Z$_1$-bond \\
		{\tiny
			$\begin{pmatrix} 
				\begin{array}{ccccc}
					-0.0183 & -0.0150 & -0.1032 & ~~0.0291 & -0.0521 \\
                    -0.0150 & ~~0.0653 & -0.0670 & -0.0074&  ~~0.1817 \\
                    -0.1032 & -0.0670 & -0.0762 & ~~0.1708 & ~~0.0028 \\
                    ~~0.0291 & -0.0074 & ~~0.1708 & ~~0.0097 & ~~0.0405 \\
                    -0.0521 & ~~0.1817 & ~~0.0028 & ~~0.0405 & ~~0.0151 
				\end{array}
			\end{pmatrix}$}
		&
		{\tiny
			$\begin{pmatrix}
				\begin{array}{ccccc}
					~~0.0323 & -0.0899 & -0.0070&  ~~0.0027 & ~~0.0193 \\
                   -0.0899 & -0.0784 & ~~0.0121 & -0.1796 & -0.0473 \\
                   -0.0070 & ~~0.0121  & ~~0.0100  &-0.0476  & ~~0.1845 \\
                    ~~0.0027 & -0.1796 & -0.0476 &  -0.0431 &  -0.0016 \\
                    ~~0.0193 & -0.0473 & ~~0.1845 & -0.0016 & ~~0.0742 
				\end{array}
			\end{pmatrix}$
		}
		&
		{\tiny
			$\begin{pmatrix}
				\begin{array}{ccccc}
					-0.0265 &  ~~0.0477 &  -0.0163&  -0.0228 & ~~0.2041 \\
                    ~~0.0477  & ~~0.0212  & ~~0.1853  &~~0.0196 & ~~0.0170 \\
                    -0.0163 &  ~~0.1853 &  ~~0.0650 & ~~0.0179   &-0.0533 \\
                    -0.0228 &  ~~0.0196 &  ~~0.0179 & ~~0.0141   & ~~0.0077 \\
                    ~~0.2041  & ~~0.0170  & -0.0533 & ~~0.0077   &-0.0782 
				\end{array}
			\end{pmatrix}$
		}
		\\
	\end{tabular}
	\begin{tabular}{ccc}
		X$_2$-bond & Y$_2$-bond & Z$_2$-bond \\
		{\tiny
			$\begin{pmatrix} 
				\begin{array}{ccccc}
					-0.0213 & ~~0.0404  &-0.0290 & ~~0.0126  &-0.0045 \\
                     ~~0.0500 & ~~0.0033  &-0.0083 & -0.0206 & -0.0514 \\
                    -0.0590 & -0.0213 & ~~0.0000 &  ~~0.0746 & ~~0.0163 \\
                     ~~0.0006 & -0.0201 & ~~0.0977 & ~~0.0323  &-0.0414 \\
                    -0.0125 & -0.0479 & ~~0.0077 & -0.0362 & -0.0100 
				\end{array}
			\end{pmatrix}$}
		&
		{\tiny
			$\begin{pmatrix}
				\begin{array}{ccccc}
					~~0.0230  &-0.0353 & ~~0.0376  & ~~0.0151  &-0.0054 \\ 
                    -0.0712 & -0.0011&  ~~0.0109 & -0.0742 &  -0.0063 \\
                    ~~0.0384  &~~0.0196  &-0.0104  & ~~0.0145  &-0.0531 \\
                    ~~0.0279  &-0.0885 & ~~0.0043  & -0.0107 & ~~0.0570 \\
                    -0.0010 & -0.0181 &  -0.0486 &  ~~0.0486 &  -0.0034 
				\end{array}
			\end{pmatrix}$
		}
		&
		{\tiny
			$\begin{pmatrix}
				\begin{array}{ccccc}
					~~0.0181 & -0.0222 & -0.0433 & -0.0306 & ~~0.0921 \\
                    -0.0316&  -0.0155&  -0.0534 &  ~~0.0241 & ~~0.0186 \\
                    -0.0386&  -0.0491&  ~~0.0015 & -0.0277 & -0.0185 \\
                    -0.0189&  ~~0.0285 & -0.0363 & ~~0.0006  &~~0.0140 \\
                    ~~0.0979 & ~~0.0110  &-0.0065  &-0.0239  &-0.0004 \\
				\end{array}
			\end{pmatrix}$
		}
		\\
	\end{tabular}
	\begin{tabular}{ccc}
		X$_3$-bond & Y$_3$-bond & Z$_3$-bond \\
		{\tiny
			$\begin{pmatrix} 
				\begin{array}{ccccc}
					~~0.0239 & ~~0.0031  &-0.0241 & ~~0.0315  &~~0.0131 \\
                    ~~0.0031 & ~~0.0086  &~~0.0073  &-0.0044  &-0.0108 \\
                    -0.0241&  ~~0.0073 & -0.0289&  ~~0.0149 & ~~0.0038 \\
                    ~~0.0315 & -0.0040 & ~~0.0149 & -0.0039 & -0.0062 \\
                    ~~0.0131 & -0.0108 & ~~0.0038 & -0.0062 & ~~0.0103 
				\end{array}
			\end{pmatrix}$}
		&
		{\tiny
			$\begin{pmatrix}
				\begin{array}{ccccc}
					~~0.0304 & ~~0.0003  &-0.0019  &-0.0269 & ~~0.0035 \\
                    ~~0.0003 & -0.0285 & ~~0.0049  &-0.0284 & ~~0.0090 \\
                    -0.0019&  ~~0.0049 & ~~0.0104  &~~0.0155  &-0.0120 \\
                    -0.0269&  -0.0288&  ~~0.0155 & -0.0105&  ~~0.0066 \\
                    ~~0.0035 & ~~0.0090  &-0.0120  &~~0.0066  &~~0.0073 
				\end{array}
			\end{pmatrix}$
		}
		&
		{\tiny
			$\begin{pmatrix}
				\begin{array}{ccccc}
					-0.0230 & -0.0140&  -0.0071 & -0.0032&  ~~0.0241 \\
                    -0.0140 & ~~0.0092 & -0.0116  &-0.0095 & ~~0.0054 \\
                    -0.0071 & -0.0116&  ~~0.0081  &-0.0009 & ~~0.0089 \\
                    -0.0032 & -0.0095&  -0.0009 & ~~0.0448 & ~~0.0010 \\
                     ~~0.0241  &~~0.0054  &~~0.0089    & ~~0.0100 & -0.0292
				\end{array}
			\end{pmatrix}$
		}
		\\
	\end{tabular}

	\label{tab:hopp}
\end{table*}

\subsubsection{Estimation of SOC strength ($\lambda$)}
\begin{figure}[ht]
	\centering
	\includegraphics[width=7 cm]{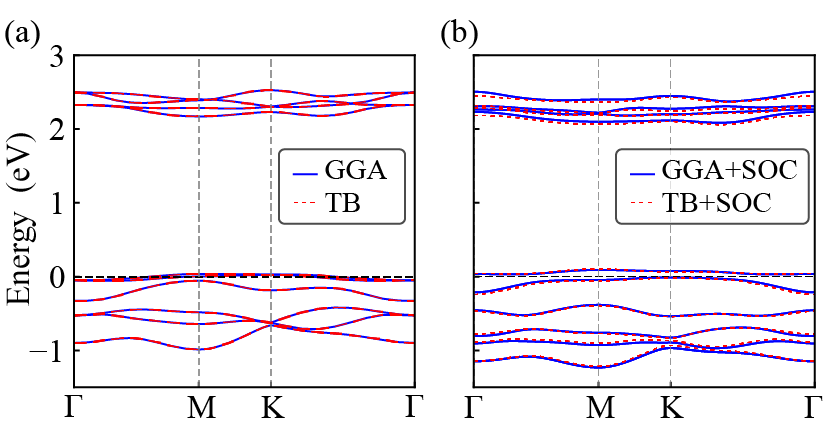}
	\caption{Comparison of \abo{} and TB model band structures for, (a) non-spin-polarized case, and (b) SOC included case. 
		In (b), SOC effects in \abo{} calculations are included self-consistently, which then fitted with the TB Hamiltonian obtained 
		in (a) after explicitly adding onsite SOC term $H_\text{soc} = \sum_i \lambda \bm L_i \cdot \bm s_i$ to $H_\text{TB}$.}
	\label{fig:socfit}
\end{figure}

To extract SOC strength in \os, we added an onsite SOC term to the TB Hamiltonian as\cite{gu2022tbsoc},
$$
H= \mathcal{I}_2 \otimes H_\text{TB} + H_{\text{soc}}
$$ 
where $\mathcal{I}_2$ is the two-dimensional identity matrix and $\otimes$ is the Kronecker product. 
$H_\text{soc} = \sum_i \lambda \bm L_i \cdot \bm s_i$ is the onsite SOC interaction term that couples the  electronic spin ($\bm{S}$) and its orbital momentum ($\bm{L}$) with a  strength $\lambda$.  Here, the SOC strength $\lambda$ is estimated by fitting the eigenvalues of $H$ with those obtained from SOC included \abo{} band structure calculations as the target. The best agreement between these two band structures, shown in Fig.~\ref{fig:socfit}(b), 
is obtained for $\lambda$ = 0.335 eV which is considered in further calculations.

\subsubsection{Estimation of Hubbard $U$ and Hund's $J_\text{H}$ within cRPA}
We estimate the Coulomb matrix elements $U_{ijkl}$($\omega$ = 0) within the constrained random phase approximation (cRPA)
implemented with VASP~\cite{crpa1,*crpa2,*crpa3}.
We neglect the screening effects for all the five Os $d$ orbital states which are energetically well-separated from other states (see Fig.~\ref{fig:socfit}).  The estimated parameters are $U$ = 3.8 eV and $J_\text{H}$ = 0.143 eV.

\subsection{Validity of \jef~=  1/2 picture}
\begin{figure*}[ht]
	\centering
	\includegraphics[width=14.0 cm]{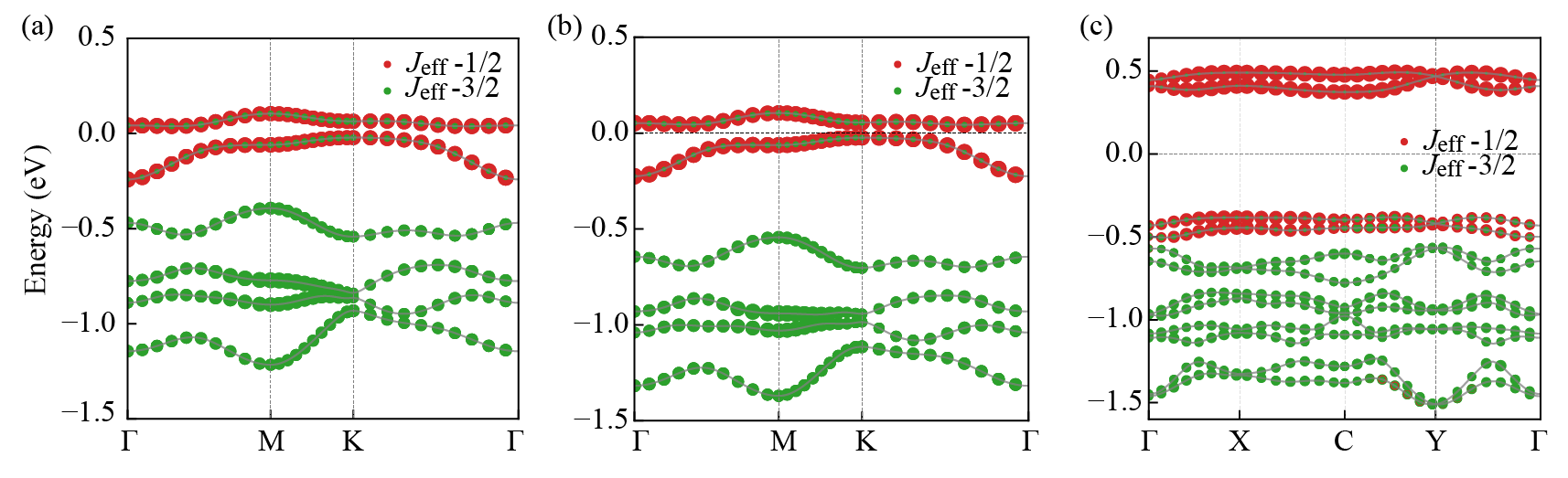}
	\caption{Band structure plots of monolayer \os~for, (a) SOC included, (b) SOC + $U$, and (c) SOC + $U$ + ZZ magnetic state are shown. Red and green spheres depicted the weight of \jef~=1/2 and 3/2 states, respectively. 
		Fermi energy is set to zero in these plots. }
	\label{fig:jeffproj}
\end{figure*}

It is crucial to analyze the effects of various electronic interactions on the band structure of monolayer \os. Such analysis can help us to examine the validity of low-energy \jef~= 1/2 picture, which is considered a building block for the emergence of dominant Kitaev interaction. To this end, \abo~band structure calculations are performed for three cases, $viz$ SOC, SOC + $U$, and SOC + $U$ + ZZ magnetic state. Obtained results are then projected onto \jef~= 1/2 and 3/2 states, eigenfunctions of which are given below in Eq.~\ref{eqn1}.  

%\begin{eqnarray} 
\begin{equation}\footnotesize
	\begin{aligned}
		\Ket{\frac{1}{2}, \pm \frac{1}{2}} 
		& =  \frac{1}{\sqrt{3}}\left( \mp \ket{d_{xy}; \pm \frac{1}{2}} \mp i \ket{d_{xz}; \mp \frac{1}{2}} - \ket{d_{yz}; \mp \frac{1}{2}} \right) \\ 
		\Ket{\frac{3}{2}, \pm \frac{3}{2}} 
		& =  \frac{1}{\sqrt{2}}\left( -i \ket{d_{xz}; \pm \frac{1}{2}} \mp \ket{d_{yz}; \pm \frac{1}{2}}  \right) \\ 
		\Ket{\frac{3}{2}, \pm \frac{1}{2}} 
		& = \frac{1}{\sqrt{6}}\left( 2 \ket{d_{xy}; \pm \frac{1}{2}} -i\ket{d_{xz}; \mp \frac{1}{2}} \mp \ket{d_{yz}; \mp \frac{1}{2}} \right)
	\end{aligned}
	\label{eqn1}
\end{equation}
%\end{eqnarray}
Plots obtained for the three cases are shown in Fig.~\ref{fig:jeffproj}(a), (b) and (c) respectively. SOC effect is included at the self-consistent level in all these calculations.  

Certain features of the band structures can immediately be identified. Firstly, one can find a clear separation between narrow \jef~= 1/2 and 3/2 bands for SOC-only band structure shown in Fig.~\ref{fig:jeffproj}(a). In Fig.~\ref{fig:jeffproj}(b), for the SOC + U case, the \jef~= 1/2 bands near the Fermi level remain largely unaffected at this $U=1$ eV (though a larger $U$ = 2 eV 
opens a clear band gap which is not shown here) and the system is nearly insulating in both cases. 
This differs from the \ru~picture presented by Kim {\it et. al.}~\cite{rucl3_bnds} where $U$ = 1.5 eV was necessary to bring out 
a clear \jef~= 1/2-3/2 separation and to drive the system insulating. The weak response towards smaller $U$ along with the narrow bandwidth ($\sim$ 0.3 eV) of \jef~= 1/2 states may be hinting that $t$ and $U$ parameters have similar scales in this material. A constrained random phase approximation 
(cRPA) estimation brings $U$ = 3.8 eV and $J_\text{H}$ = 0.143 eV for monolayer \os. However, the largest  
1NN hopping magnitude $|t|$ estimated from the Wannier TB model is $\approx$ 0.2 eV. 
Based on these values, one can still think of limit $t \ll U$ and hence the formation of 
well-localized \jef~states in this compound. A wider band gap opens up once magnetism in the form of ZZ state is included and obtained band structure for this case is shown in Fig.~\ref{fig:jeffproj}(c).

%$|${\bf a}$|$ = 12.355, $|${\bf b}$|$ = 6.188 \AA{}, with vacuum $|${\bf c}$|$ = 25.0 \AA{} and $\alpha$ = $\beta$ = 90$^\circ$ ~and $\gamma$ = 109$^\circ$ $|${\bf a}$|$ = 12.355, $|${\bf b}$|$ = 6.188 \AA{}, with vacuum $|${\bf c}$|$ = 25.0 \AA{} and $\alpha$ = $\beta$ = 90$^\circ$~and $\gamma$ = 109$^\circ$. \
%Using plane wave cutoff energy 500 eV, $4\times8\times1$ $\Gamma$-centered $k$-mesh and energy convergence criteria of 10$^{-5}$ eV, we optimize the lattice parameters with experimentally proposed magnetic ground  % considering SOC effect at the self-consistent level. A DFT+$U$ approach employing Liechtenstein~\cite{liech} scheme with
%on-site Coulomb interaction $U$ = 2.5 eV and exchange interaction $J_\text{H}$ = 0.9 eV was used. The values of $U$ and
% $J_\text{H}$ parameters are consistent with the previous study~\cite{calder_sro}. Optimized $a$ and $b$ lattice constants were foundto be overestimated by $\sim$ 3.1 \% while $c$ remains the same. Since this change in lattice constants is significant, we have used the optimized structure in further calculations.

\subsection{Second-order perturbation method}
The full Hamiltonian for the $d^5$ Os$^{+3}$ ions is given by, 
\begin{equation}
	\mathcal{H} = H_{\text{hop}} + H_{\text{cf}} + H_{\text{soc}} + H_{\text{int}}
\end{equation}
where the terms on the right-hand side are hopping, CF, SOC, and Hubbard-Kanamori interaction Hamiltonians. 
Using the basis $\psi^{\dagger}_{i\sigma}$ = $[d^\dagger_{z^2}$,$d^\dagger_{xz}$, $d^\dagger_{yz}$, $d^\dagger_{x^2-y^2}$, $d^\dagger_{xy}]_{i\sigma}$ at site $i$ of Os and spin $\sigma$, above Hamiltonian then reads, 
\begin{equation}\footnotesize
	\begin{aligned}
	\mathcal{H} &=  \sum_{i\neq j,\sigma} \psi_{i\sigma}^\dagger T_{ij}\psi_{j\sigma} + \sum_{i,\sigma} \psi_{i\sigma}^\dagger {\Delta^\text{CF}_i}\psi_{i\sigma} + \sum_i \lambda \bm L_i \cdot \bm s_i \\
	&+ \frac{U}{2} \displaystyle\sum_{i,\alpha} n_{i\alpha \sigma} n_{i\alpha \sigma'} 
	+ \frac{U'}{2} \displaystyle\sum_{i,\alpha \ne \beta} n_{i \alpha}n_{i \beta}  \\ 
	&-\frac{J_\text{H}}{2} \displaystyle\sum_{i, \sigma, \sigma', \alpha \ne \beta} \psi^\dagger_{i\alpha \sigma} \psi_{i\alpha \sigma'} \psi^\dagger_{i\beta \sigma'} \psi_{i\beta \sigma} \\
	&-\frac{J'}{2} \displaystyle\sum_{i, \sigma \ne \sigma', \alpha \ne \beta} \psi^\dagger_{i\alpha \sigma} \psi_{i\beta \sigma'} \psi^\dagger_{i\alpha \sigma'} \psi_{i\beta \sigma}
	\end{aligned}
	\label{eq:full}
\end{equation}

In the above expression, the first three terms are hoppings, CF and SOC, while the last four terms are the Hubbard-Kanamori interactions. The $U$/$U'$ are
intraorbital/interorbital Hartree energies and  $J_\text{H}$ and $J'$ are Hund's coupling
and pair hopping interaction,
respectively. Rotational invariance in the isolated atom limit dictates the relationships: $U'$ = $U$ - 2$J_\text{H}$ and $J_{\text H}$ = $J'$. Hopping amplitudes $ T_{ij}$ and CF matrix $\Delta^\text{CF}_i$ can be obtained by fitting 
the \abo{} band structure with a TB model as described earlier. 

In the limit $t \ll U$, i.e. in the isolated atom limit, one can separate the ``onsite'' term $H^0_i$ at a site $i$ from $\mathcal{H}$ as, 
$$
H^0_i = H_{\text{cf}} + H_{\text{soc}} + H_{\text{int}}
$$
We drop index $i$ since this Hamiltonian is the same for each site.
The key here is that the $d^5$ manifold has a two-fold degenerate ground state, 
which forms a  Kramers doublet and can be treated as a pseudospin-1/2.  In order to extract the magnetic 
interactions of these pseudospin states,  we first project the full TB Hamiltonian onto the pseudospin $j_{1/2}$ 
space $\{\phi_{i\alpha}\}$, $\alpha = \uparrow, \downarrow$, where $\uparrow,\downarrow$ 
refer to the SOC pseudospin-1/2 states. Starting from the isolated limit we introduce $H_\text{hop}$ as a perturbation.
In the second-order perturbation theory, 
the Hamiltonian is written as,

\begin{eqnarray}
	H^{(2)} &=& \sum_{ij}\sum_{\alpha\beta\alpha'\beta'} \mathcal H (i,j)_{\alpha\beta\alpha'\beta'}
	| i\alpha,j\beta\rangle\, \langle i\alpha', j\beta '|, \nonumber
\end{eqnarray}

\begin{equation}
	\begin{aligned}
	\mathcal H (i,j)_{\alpha\beta\alpha'\beta'} = \sum_{kl}\sum_{\gamma\lambda}
	 & \frac{1 }{\Delta E} 
	\langle i\alpha,j\beta|H_{\text{hop}}|k\gamma, l\lambda\rangle   \\
 	& \times \langle k\gamma,l\lambda|H_{\text{hop}}|i\alpha', j\beta'\rangle ,
	\end{aligned}
	\label{eq:h02}
\end{equation}

where $1/\Delta E = \frac{1}{2}[{1}/(E_{i\alpha}+E_{j\beta} - E_{k\lambda}-E_{l\gamma}) +
{1}/(E_{i\alpha'}+E_{j\beta'} - E_{k\lambda}-E_{l\gamma})]$.
Here, $| i\alpha,j\beta\rangle$ and $| i\alpha', j\beta'\rangle$ are two-site states in the \jef~= 1/2 ground states, and $| k\lambda, l\gamma\rangle$ are two-site excited states, 
both in the isolated atom limit.  $H_{\text{hop}}$ connects a two-site ground state to an excited state with $(d^4, d^6)$ configuration, the dimensions of whose Hilbert spaces 
is 210.  The eigenstates of isolated Os with 4, 5, and 6 $d$ electrons are obtained by exact diagonalization. 
One can represent the pseudo-spins $j_{1/2}$ as $S^{\mu} = \braket{i\alpha| \bm{j}^\mu_{i,\text{eff}}|i\beta}$
which are the expectation values of pseudospin $\bm{j}^\mu_\text{eff}$ operators with $\mu = 0,x,y,z$. 
Here, $\bm{j}^0_\text{eff}$ = $\mathcal{I}_2$ is the matrix representation of operator 
$\bm{j}^0_\text{eff}$. 
Eq.~(\ref{eq:h02}) can be mapped to a spin Hamiltonian of the form, 
\begin{equation}
	\begin{aligned}
	H_{\text{spin}} &= S_i^\mu \eta(i,j)^{\mu\nu} S_j^\nu  \\
	&= \frac{1}{4} 
	\eta(i,j)^{\mu\nu}  \phi _{i\alpha }^{\dagger }\sigma _{\alpha\alpha '}^\mu \phi _{i \alpha '} 
	\phi _{j\beta } \sigma _{\beta\beta '}^{\nu}  \phi _{j\beta }^{\dagger },
	\end{aligned}
\end{equation}
where $\mu,\nu=0,x,y,z$, $\sigma^\mu$ are Pauli matrices, and summation over all repeated indexes is implied. The map can be achieved by solving the linear equations
\begin{equation}
	-\frac{1}{4} \sigma_{\alpha\alpha'}^\mu \sigma_{\beta\beta'}^\nu 
	\eta(ij)^{\mu\nu}
	=\mathcal H(i,j)_{\alpha\beta\alpha'\beta'}
\end{equation}

Thus, the most general form of exchange interaction matrix on the three Os-Os bonds can be written as,  
\begin{equation}
	\begin{gathered}
\eta_\text{X} = 
\begin{pmatrix}
	J + K & & \Gamma' + \zeta & & \Gamma' - \zeta \\
	\Gamma' + \zeta && J + \xi && \Gamma \\
	\Gamma' - \zeta && \Gamma && J - \xi \\
\end{pmatrix}\\
\eta_\text{Y} = 
\begin{pmatrix}
	J + \xi && \Gamma' + \zeta && \Gamma \\
	\Gamma' + \zeta && J + K && \Gamma' - \zeta \\
	\Gamma && \Gamma' - \zeta && J - \xi \\
\end{pmatrix}\\
\eta_\text{Z} = 
\begin{pmatrix}
	J + \xi && \Gamma  && \Gamma' + \zeta \\
	\Gamma && J - \xi && \Gamma' - \zeta \\
	\Gamma' + \zeta && \Gamma' - \zeta && J + K \\
\end{pmatrix}
	\end{gathered}
	\label{operation}
\end{equation}

Only these symmetric components of interaction matrices appear on the first and third neighbor bonds, while on the second neighbor bonds, due to the absence of inversion symmetry, an additional antisymmetric component appears in the form of Dzyaloshinskii–Moriya interaction (DMI). 
The matrix from these antisymmetric interactions is given in Ref.~\cite{rucl3_ref}.  

\subsubsection{Three and four orbital terms}
Since we consider both $t_{2g}$ and $e_g$ orbitals in our basis above, to ensure the rotational invariance of the ``onsite'' Hamiltonian $H^0_i$, one should include three and four orbital terms in the Hubbard-Kanamori Hamiltonian. Three orbital terms can be given, 
\begin{equation}\footnotesize
	\begin{aligned}
	H^\text{3orb}_\text{int} & = \frac{V_1}{2} \displaystyle\sum_{ \sigma, \sigma', \alpha \ne \beta} \psi^\dagger_{i\alpha \sigma} \psi_{i\beta \sigma} \psi^\dagger_{i\gamma \sigma'} \psi_{i\gamma \sigma'} \\
 &+ \frac{V_1}{2} \displaystyle\sum_{ \sigma, \sigma', \alpha \ne \beta} \psi^\dagger_{i\alpha \sigma} \psi_{i\gamma \sigma} \psi^\dagger_{i\gamma \sigma'} \psi_{i\beta \sigma'} \\
	&+\frac{V_2}{2} \displaystyle\sum_{ \sigma \ne \sigma', \alpha \ne \beta} \psi^\dagger_{i\alpha \sigma} \psi_{i\gamma \sigma} \psi^\dagger_{i\beta \sigma'} \psi_{i\gamma \sigma'}\\
 &+\frac{V_2}{2} \displaystyle\sum_{ \sigma \ne \sigma', \alpha \ne \beta} \psi^\dagger_{i\gamma \sigma} \psi_{i\alpha \sigma} \psi^\dagger_{i\gamma \sigma'} \psi_{i\beta \sigma'}
	\end{aligned}
	\label{eq:three_orb}
\end{equation}
In the above equation, indices $\alpha$, $\beta$ = $d_{z^2}$/$d_{x^2-y^2}$ and $\gamma$  = $d_{xz}$/$d_{yz}$. Magnitude of $V_1$, $V_2$ are usually quite small ($<$ 10$^{-2}$ eV). Four orbital terms in the Hubbard-Kanamori Hamiltonian can be given as, 
\begin{equation}\footnotesize
	\begin{aligned}
	H^\text{4orb}_\text{int} & = \frac{V}{2} \displaystyle\sum_{ \sigma, \sigma', \alpha \ne \beta} \psi^\dagger_{i\alpha \sigma} \psi_{i\delta \sigma} \psi^\dagger_{i\beta \sigma'} \psi_{i\gamma \sigma'} 
	\end{aligned}
	\label{eq:four_orb}
\end{equation}
In the above equation, the four distinct orbital indices $\alpha$, $\beta$, $\gamma$, and $\delta$ run over all the five $d$ orbitals in our basis. The magnitude of $V$ is usually of the order of $\sim$ 10$^{-3}$ eV. Addition of $H^\text{3orb}_\text{int}$ and $H^\text{4orb}_\text{int}$ to $H_\text{int}$ term of Eq. 2 in the main text modify the magnetic interactions mostly at the third decimal places. This is consistent with the fact that these additional terms are redundant in the presence of large crystal field splitting($\Delta^\text{CF}_i$ in Eq.~\ref{eq:cf}) which is also the case of monolayer \os.
Since these significantly smaller changes in magnetic interactions do not bring any substantial changes in the magnetic ground state, we used only two orbital terms of Eq. 2 for the phase diagram in Fig. 2(a) of the main text.

\subsection{Pseudo-fermion functional renormalization group calculations}
We use  SpinParser package~\cite{SpinParser1} to perform pseudofermion functional renormalization group (pf-FRG) calculations~\cite{pfrg1,pfrg2} on generalized Kitaev spin model obtained in the previous section. 
We consider honeycomb lattice to initialize vertex functions in order to capture two-particle interactions on lattice sites separated by up to 7 lattice bonds. In pf-FRG calculations, the single-particle vertex $ \Sigma^{\Lambda}(\omega) $ and two-particle vertex basis functions $ \Gamma^{\Lambda}(\omega, \omega^\prime, \omega^{\prime\prime})$ are parameterized by a frequencies $\omega$ and the renormalization group cutoff $\Lambda$. 
Here, then we use a logarithmically spaced mesh of discrete frequencies as 
$$ 
\omega_{n}=\omega_{\min }\left(\frac{\omega_{\max }}{\omega_{\min }}\right)^{\frac{n}{N_{\omega}-1}};\quad n=(0,1,\cdots,N_\omega-1).
$$
where $\omega_{\min}$ and $\omega_{max}$ are the limit points of frequency on the positive half-axis which are set to be 0.01 and 35.0 respectively. $N_{\omega}$ is the overall number of positive frequencies and is chosen to be 32 in our case. As for the negative half-axis, the  frequencies are generated automatically by symmetry. The discretization of the cutoff parameter $\Lambda$ is also done on a logarithmically spaced mesh given by,
$$
\Lambda_{n}=\Lambda_{\max } b^{n}; \quad   n=0, \ldots,\left\lfloor\log _{b}\left(\frac{\Lambda_{\min }}{\Lambda_{\max }}\right)\right\rfloor 
$$
where the $\Lambda_{\min}$ and $\Lambda_{\max}$ are the lower and upper boundaries of $\Lambda$ which are set to be 0.01 and 50, respectively. The parameter $b$ is step size and is set to be 0.98 in our calculations. 
Though we have considered up to 3NN interactions in our pf-FRG calculations, negligibly small 2NN and 3NN magnetic interactions do not affect the magnetic ground state obtained with this method at a qualitative level. We find that most part of the phase points in Fig. 2(a) of the main text can be described by dominant 1NN magnetic interactions only.

%ion for the general fermionic functional renormalization group framework. It maps the spin Hamiltonian onto a a (pseudo)fermionic model and then analyzes it using the fermionic FRG framework. Here we adopt the SpinParser package ~\cite{SpinParser1} to perform the pf-FRG analysis on the 
%r OsCl$_3$, the lattice type is set to be honeycomb for the SpinParse to initial vertex functions to capture two-particle interactions on lattice sites up to a truncation of 7 lattice bonds apart. 
%^{\Lambda}(\omega, \omega^\prime, \omega^{\prime\prime})$ are parameterized by a frequencies $\omega$ and the renormalization group cutoff $\Lambda$. 
%
%
%cdots,N_\omega-1).
%
%t to be 0.01 and 35.0 respectively. $N_{\omega}$ is the overall number of positive frequencies and chosen to be 32 in our case. As for the negative half-axis, the  frequencies are generated automatically by symmetry. The discretization of the cutoff parameter $\Lambda$ is also used a logarithmically spaced mesh, as
%
%max }}\right)\right\rfloor 
%
%0.01 and 50, respectively. The parameter $b$ is step size and set to be 0.98 in our calculations.
%<frequency discretization="exponential">
%	<min>0.01</min>
%	<max>35.0</max>
%	<count>32</count>
%</frequency>
%<cutoff discretization="exponential">
%	<max>50.0</max>
%	<min>0.01</min>
%	<step>0.98</step>
%</cutoff>
%<lattice name="honeycomb" range="7"/>

\subsection{Atomic features}
One of the quantities which can be measured from the resonant
inelastic X-ray scattering (RIXS) experiments are single-point excitations represented by sharp peaks in the
scattering intensity in the relevant energy range.  
It can be a direct probe for cubic symmetry lowering of the Rh-O$_6$ octahedra in a material.
Theoretically, such a low-lying crystal field-assisted many-body excitations bear a  
close resemblance with the eigenvalues obtained from diagonalization of many-body
``onsite'' Hamiltonian $H^0$. Hence, we discuss here the nature of CF splitting in monolayer \os~and its effect 
on the electronic structure. The CF matrix is given in Eq.~\ref{eq:cf}. 

Examining eigenvalues of CF matrix ($\Delta_i$) in Eq.~\ref{eq:cf} reveals that $t_{2g}$-$e_g$ 
$\Delta_i^{\tiny t_{2g}-e_g}$ $\approx$ 2.60 eV and there is additional trigonal distortion 
$\Delta_i^\text{\tiny tri}$ $\approx$ 64 meV of the octahedra. Eigenvalues obtained after exact diagonalization of $H_i^0$ with this $\Delta^\text{CF}_i$ can be compared with the single-point excitations observed in RIXS experiments. In our calculation, we find \jef~=1/2 - 3/2 separation to be 0.580 eV which is approximately 
$\frac{3}{2}\lambda$ for $\lambda$ = 0.355 eV estimated from band structure fitting of monolayer \os. 
The four-fold \jef~=3/2 states are split into two doublets which are separated by $\approx$ 56 meV. This splitting is 
a direct consequence of $\Delta_i^\text{\tiny tri}$ present and is also found in Iridates and \ru. By 
putting $\Delta_i^\text{\tiny tri}$ = 0 eV, as expected, \jef~=3/2 states regained the four-fold degeneracy.

\subsection{Magnetic interaction and ground state from cRPA parameters}
Below in Table~\ref{tab:crpa_int}, we provide estimated magnetic interactions corresponding to parameters obtained from cRPA, 
i.e. $U$ = 3.8 eV and $J_\text{H}$ = 0.143 eV.

\begin{table}[!ht]
	\centering
	\caption{ Estimated bond-dependent magnetic interactions for cRPA estimated $U$ = 3.8 eV, $J_{\text{H}}$ = 0.143 eV. Here also, $\lambda$ = 0.355 eV.  Values of  Heisenberg $J$, Kitaev $K$, 
		off-diagonal $\Gamma$, and $\Gamma'$  terms along with diagonal and off-diagonal anisotropic terms $\xi$ and $\zeta$ are listed. 3NN interactions as well as some 2NN 
		interactions are $<$ 0.05 meV and hence not listed here.}
	\begin{tabular}{ccccccc}
		\hline \hline
		%\specialrule{0em}{0pt}{2pt}
		% &  & \multicolumn{3}{c}{Set 1}  &   \\
		%\specialrule{0em}{0pt}{2pt}
		%\cline{2-7}
		%\specialrule{0em}{0pt}{2pt}
		%  Bond             & $J$        & $K$     & $\Gamma$  & $\Gamma'$  & $\xi$  & $\zeta$  \\
		% \hline 
		%\specialrule{0em}{0pt}{3pt}
		%X$_1$              &   ~~0.133    & -1.824  &   -0.445  & ~0.078     & ~0.104 & -0.174 \\ 
		%Y$_1$              &   ~~0.131    & -1.753  &   -0.449  & -0.104     & -0.062 & -0.228 \\  
		%Z$_1$              &   ~~0.100    & -1.755  &   ~0.410  & ~0.308     & -0.100 & ~0.040 \\ 
		%X$_2$/Y$_2$/Z$_2$  &   ~-0.211    & ~0.630  &     --    &  --        &   --   &  --     \\
		%\specialrule{0em}{0pt}{3pt}
		%\cline{2-7} 
		%\specialrule{0em}{0pt}{2pt}
		%&  & \multicolumn{3}{c}{$\bm{D}_{ij}$}  &   \\
		%\hline
		%\specialrule{0em}{0pt}{3pt}
		%X$_2$           &   &   (-0.022, & -0.044, &  ~0.041)  &   &  \\ 
		%Y$_2$           &   &   ( 0.043, & -0.044, & -0.019)  &  &  \\  
		%Z$_2$           &   &   ( 0.039, & -0.016, & ~0.047)   &  & \\
		%\specialrule{0em}{0pt}{3pt}
		%%\cline{2-7}
		%%X$_3$/Y$_3$/Z$_3$ & &    &   &   $\le$ $|$0.05$|$  &   &   &     \\ \\
		%\hline \hline
		\specialrule{0em}{0pt}{2pt}
		& &\multicolumn{4}{c}{Magnetic interactions (meV)}  &  \\
		\specialrule{0em}{0pt}{2pt}
		\cline{2-7}
		\specialrule{0em}{0pt}{2pt}
		Bond   & $J$   & $K$ & $\Gamma$ & $\Gamma'$  & $\xi$  & $\zeta$  \\
		\hline 
		\specialrule{0em}{0pt}{3pt}
		X$_1$ &  ~0.057    & -1.318  &  ~0.428  &  -0.334 &  -- & 0.235    \\ 
		Y$_1$ &  ~0.068    & -1.571  &  ~0.465  & -0.051 & -- & -0.364 \\  
		Z$_1$ &  ~0.043    & -1.396  &   ~0.372  & -0.212 & -- &  -0.388 \\
		%X$_2$ & -0.211    &   0.573  &  -0.137  & -0.056 &  0.042 & -0.088     \\
		X$_2$ & --     &   ~~0.075  &  --  & -- &  -- & --     \\
		Y$_2$ & ~0.132     &   ~~0.126  &  --  &  -- & -- & --     \\
		
		%Z$_2$ & -0.234    &   0.636  &  0.115  &  0.074 & -0.043 & -0.167     \\
		Z$_2$ & --    &  ~~0.068  &  --  &  -- & -- & --     \\
		
		\specialrule{0em}{0pt}{2pt}
		\cline{2-7} 
		\specialrule{0em}{0pt}{2pt}
		&  &   \multicolumn{3}{c}{$\bm{D}_{ij}$} &    \\
		\hline
		\specialrule{0em}{0pt}{3pt}
		X$_2$           &   & &      &  \\ 
		Y$_2$           &   & &  $<$ $|$0.04$|$   &  \\  
		Z$_2$           &   & &      & \\
		\specialrule{0em}{0pt}{2pt}
		\hline \hline
	\end{tabular}
	\label{tab:crpa_int}
\end{table}

\begin{figure}[ht]
	\centering
	\includegraphics[width=7.0 cm]{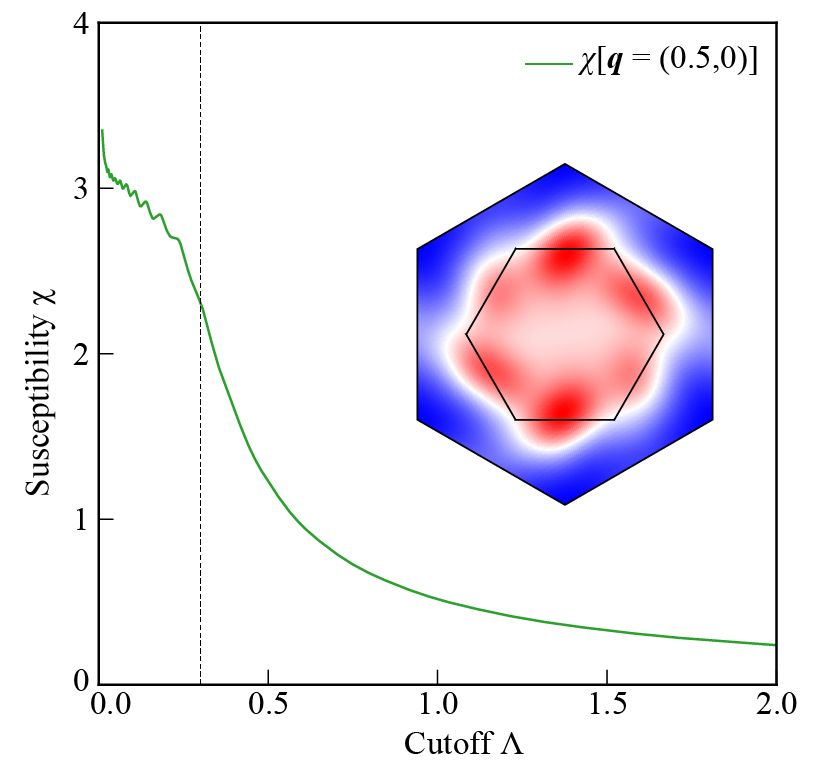}
	\caption{Static spin correlations plot, shown in the inset, obtained from pf-FRG calculations using magnetic interactions from Table~\ref{tab:crpa_int}. Intensity contribution from one of the $M$ points of the BZ indicates a ZZ ground state. The renormalization group flow of the magnetic correlation ($\chi$) represented by the green curve breaks at $\approx$ 0.3 $\Lambda$ indicating the onset of the long-range magnetic order.}
	\label{fig:crpa}
\end{figure}

We use these interactions from Table~\ref{tab:crpa_int} in the pf-FRG calculations and the static spin correlation for the corresponding ZZ ground state is shown in Fig.~\ref{fig:crpa}. One of the $M$ points of the first BZ contributes to this ground state indicating some sort of ``anisotropic'' ZZ state.  

\subsection{Quantitatively different zigzag states in phase diagram}
\begin{figure}[ht]
	\centering
	\includegraphics[width=7.0 cm]{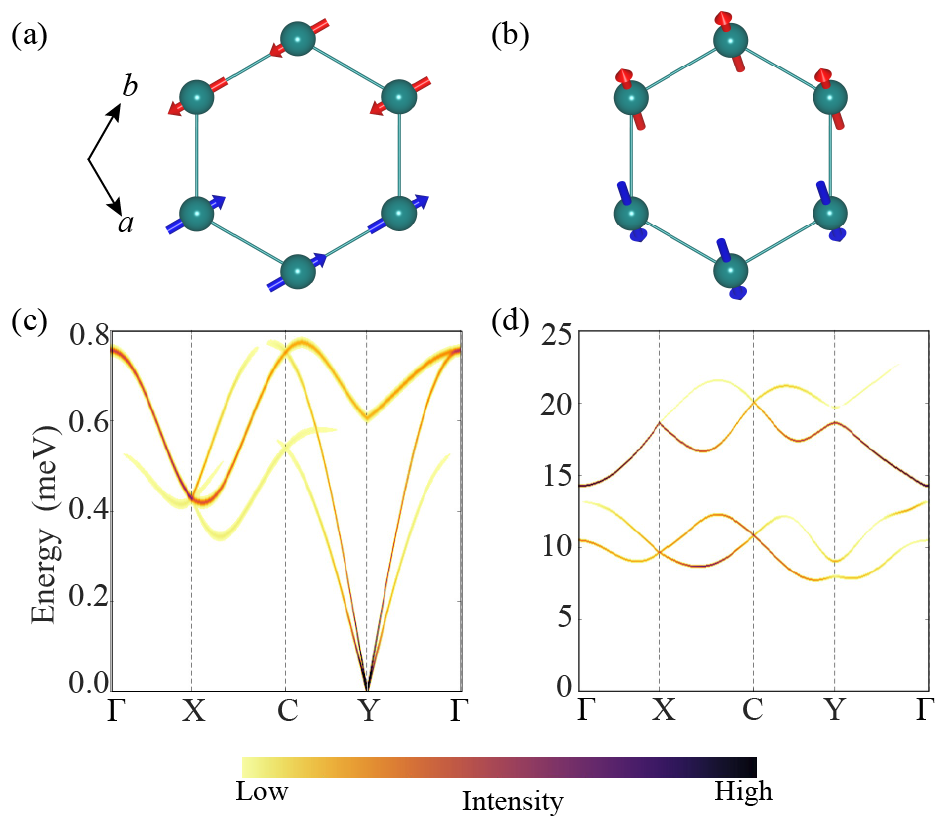}
	\caption{Optimized classical magnetic ground states for interactions corresponding to, (a) $J$ = 0.4 meV and $K$ = -18.2 meV, and (b) 
		interaction listed in Table~\ref{tab:crpa_int}. Spin wave spectra obtained using linear spin wave theory for, (c) in-plane configuration  shown in (a), and (d) out-of-plane shown in (b)}
	\label{fig:spinw}
\end{figure}

As we have mentioned that different part of the phase diagram in Fig. 2(a) of the main text hosts different magnetic ground state depending upon the nature of dominant magnetic interactions. The largest part of the phase diagram is occupied by the ZZ magnetic state which differs quantitatively in different regions of the phase space. To demonstrate this point, we plot two of them in Fig.~\ref{fig:spinw}. Spin orientations in these two ZZ states are obtained by optimization of classical ground state using SpinW package~\cite{spinw}. As one can see in Fig.~\ref{fig:spinw}(a), spin alignment is in-plane while spins are tilted in the out-of-plane direction in Fig.~\ref{fig:spinw}(b). In-plane configuration corresponds to $J$ = 0.4 meV and 
$K$ = -18.2 meV and out-of-plane configuration is obtained from interactions listed in Table~\ref{tab:crpa_int}. 
Corresponding spin-wave spectrum, calculated using linear spin-wave theory as implemented in SpinW~\cite{spinw}, are also shown in Fig.~\ref{fig:spinw}(c) and (d).

As one can see that the spectra are gapless in Fig.~\ref{fig:spinw}(c) for the in-plane alignment of spins and are gapped in Fig.~\ref{fig:spinw}(d) for the out-of-plane tilting of spins. 
This distinctive feature of the spectra can be understood as follows.  In the absence of magnetic couplings like $\Gamma$ and $\Gamma'$, any in-plane spin alignment would classically have the same energy resulting in a gapless mode. This can be seen for the spin alignment of Fig.~\ref{fig:spinw}(a) results in a gapless mode at $Y$ point. However, $\Gamma$ and $\Gamma'$ terms dictate 
the out-of-plane tilting of spins and pinning down of moments in a specific direction (along one of the 
Cl-Cl edges in Os-Cl$_6$ octahedra, see Fig. 1 (a)). This introduces additional anisotropy in  the
Hamiltonian causes the spectra to be gapped. Diagonal anisotropic term $\xi$ may lead to further 
anisotropy in the system, in-tern actively contributing to gapped spectra. 

%\bibliography{ref.bib}

\end{document}